\documentclass[aps,twocolumn]{revtex4}

\usepackage{amsmath, amsthm, amssymb}

\begin{document}

\bibliographystyle{apsrev}

\title{Computing waveforms for spinning compact binaries in quasi-eccentric orbits}

\author{Neil J. Cornish and Joey Shapiro Key}
\affiliation{Department of Physics, Montana State University, Bozeman,
MT 59717}

\begin{abstract}
Several scenarios have been proposed in which the orbits of binary black holes
enter the band of a gravitational wave detector with significant eccentricity.
To avoid missing these signals or biasing parameter estimation it is important
that we consider waveform models that account for eccentricity. The ingredients
needed to compute post-Newtonian (PN) waveforms produced by spinning black holes inspiralling on
quasi-eccentric orbits have been available for almost two decades at 2 PN order,
and this work has recently been extended to 2.5 PN order. However, the computational
cost of directly implementing these waveforms is high, requiring many steps per orbit to
evolve the system of coupled differential equations. Here we employ the standard techniques
of a separation of timescales and a generalized Keplerian parameterization of the orbits to
produce efficient waveforms describing spinning black hole binaries with arbitrary masses
and spins on quasi-eccentric orbits to 1.5 PN order. We separate the fast orbital timescale
from the slow spin-orbit precession timescale by solving for the orbital motion in
a non-interial frame of reference that follows the orbital precession. We outline a scheme for
extending our approach to higher post-Newtonian order.
\end{abstract}

\maketitle

\section{Introduction}

Standard scenarios for the formation of black hole binaries predict that the
orbits will have circularized~\cite{Krolak} by the time the system reaches the late inspiral phase probed by ground
or space based gravitational wave detectors. There are however, alternative scenarios that produce
systems with significant residual eccentricity for stellar mass systems~\cite{Wen:2002km,O'Leary:2008xt};
intermediate mass systems~\cite{AmaroSeoane:2009yr,AmaroSeoane:2009cg}; and supermassive
systems~\cite{Colpi:1999cm,Aarseth:2002ie,Armitage:2005xq,Berczik:2005ff,Dotti:2005kq,Sesana:2010qb}.
Neglecting the effects of eccentricity on the waveforms will hurt
detection~\cite{Martel:1999tm,Tessmer:2007jg,Brown:2009ng} and parameter estimation~\cite{Cutler:2007mi}.

In most scenarios the individual black holes that make up the binary will be spinning, so what is
needed are waveform templates that include the effects of spin and eccentricity. The equations
of motion describing such systems, along with expressions for the instantaneous waveforms and energy and
angular momentum fluxes were computed to 2nd post-Newtonian (PN) order (order $(v/c)^4$ in relative velocity
of the system) by Kidder, Wiseman and Will in the early 1990's~\cite{Kidder:1992fr,Kidder:1995zr},
and more recently the calculation was carried to 2.5 PN order~\cite{Faye:2006gx,Blanchet:2006gy}.
By evolving these expressions for the coupled ordinary differential equations describing the orbital
motion, spin precession and energy and angular momentum decay, it is possible to generate waveforms
that can be used for data analysis and parameter estimation. The drawback of this direct approach
is the high computational cost associated with accurately capturing the rapid orbital motion along
with the smaller and slower effects of periastron precession, spin precession and orbital decay.
A numerical implementation also loses control of the post-Newtonian expansion, introducing 
higher order effects that can lead to (potentially spurious) chaotic
behavior~\cite{Levin:1999zx,Hughes:2001gh,Cornish:2002eh}.

Our goal here is to develop an efficient approach for producing waveforms for spinning eccentric
binaries with arbitrary spins and eccentricities for use in gravitational wave astronomy~\cite{Key:2010tc}. 
Given the long history of the problem, it is surprising that such waveforms do not already exist in
the literature. Waveforms are available in the case of non-spinning eccentric binaries where analytic expressions
for the orbital motion and phase evolution are available at 3 PN and 3.5 PN order,
respectively~\cite{Konigsdorffer:2006zt,Damour:2004bz,Memmesheimer:2004cv,Gopakumar:1997bs,Arun:2007sg,Arun:2007rg,Arun:2009mc}. These can be combined with
2 PN accurate expression for the spinless waveform amplitudes~\cite{Gopakumar:2001dy} to produce ready to use
gravitational waveforms. Efficient waveforms are also known for circular binaries with spin out to
1.5 PN order for precessing systems, and out to 2 PN order for non-precessing systems~\cite{Arun:2008kb}.
For the more general case of eccentric binaries with spin, analytic expressions for the orbital motion
and phase evolution have been computed to 1.5 PN order for equal mass systems and systems with one
spinning body~\cite{Konigsdorffer:2005sc}. These can be combined with the 1.5 PN order expression for
the waveform amplitudes~\cite{Majar:2008zz} to produce ready to use waveforms for this sub-class of
systems (however we caution that the expressions in Ref.~\cite{Konigsdorffer:2005sc} include some
spurious spin-orbit effects that fail to vanish in the zero spin limit, and the waveform amplitudes quoted
in Ref.~\cite{Majar:2008zz} contain several errors. We discuss and correct both in what follows).

Here we provide, for the first time, efficient waveforms at 1.5 PN order that describe eccentric binaries
with arbitrary (but moderate) mass ratio and general spins. As is the case with the efficient waveform
models described
above, our derivation relies on the separation of timescales available in the problem that was first
described by Damour~\cite{Damour:1983tz}. Our work extends the 1 PN accurate Post-Keplerian parameterization
of the orbits introduced by Damour \& Deruelle~\cite{Damour:1985to} to include the leading order, 1.5 PN
spin-orbit effects, for both the radial and angular motion (the radial component had previously been computed
at this order by Wex~\cite{wex}). The key new element in our derivation, which allows us to analytically
compute both the radial and angular components of the post-Keplerian description of the orbital motion,
is that we re-cast the equations of motion in a non-inertial frame that follows the spin-orbit induced
precession of the orbit. This allows us to maintain the separation between the fast orbital motion
(which can be solve analytically), and the much slower spin-orbit precession (which must be solve numerically
for general masses and spins). We go on to compute the 1.5 PN order spin effects on the secular decay of
the eccentricity and semi-major axis, building on the 1 PN accurate treatment of Junker and
Sch\"{a}fer~\cite{Junker}, and revisiting the earlier calculation of Gergely, Perjes and
Vasuth~\cite{Gergely:1998sr}, who used a different spin-supplementary condition. Our expressions for
the orbital decay match - after converting from the eccentricity and semi-major axis to energy and
angular momentum - with the gauge invariant expressions for the orbital decay found earlier by
Rieth and Sch\"{a}fer~\cite{Rieth:1997mk}. We also revisit the
calculation of the waveform amplitudes~\cite{Majar:2008zz} and correct several errors in the published
expressions. Combining all these elements we arrive at ready to use, efficient, 1.5 PN order accurate
waveforms for general, spinning, eccentric binaries. We conclude with some
thoughts about continuing the calculation to higher post-Newtonian order.

\section{Equations of Motion}

We begin by deriving a semi-analytic solution to the dissipationless equations of
motion at 1.5 PN order. This we accomplish in two steps - first we find an analytic
solution to the equations of motion in a non-inertial frame that precesses with the
orbital plane. Next we calculate the time dependent rotation between the inertial
and precessing frame using a fast-slow decomposition of the spin-orbit precession
equations.

The equations of motion at 1 PN order were solved by Damour \& Deruelle~\cite{Damour:1985to} using
a generalized Keplerian parameterization of the orbits. In what follows we will
focus on the 1.5 PN order corrections, which can be added to the earlier
result to give the complete solution at this order. The 1.5 PN equations of motion
are most readily solved in the gauge defined by the Pryce-Newton-Wigner~\cite{Pryce,NW}
(PNW) spin-supplementary condition. Suppressing 1 PN terms, the relative separation of
the binary system ${\bf r}$, the individual spins ${\bf S}_{1},{\bf S}_{2}$ and the orbital angular
momentum ${\bf L}$ evolve according to the equations~\cite{wex,Konigsdorffer:2005sc}
(in units where $G=c=1$)
\begin{equation}\label{reqn}
{\bf \dot r}=\frac{{\bf p}}{\mu}+\frac{{\bf S_{\rm eff}}\times {\bf r}}{r^3}
\end{equation}
and
\begin{eqnarray}\label{so}
\frac{d {\bf L}_N}{dt} & =& \frac{1}{ r^3} {\bf S_{\rm eff}} \times {\bf L}_N \, , \\
\frac{d {\bf S}_{1}}{dt} & = & \frac{\delta_{1} }{r^3} {\bf L}_N \times {\bf S}_{1} \,, \\
\frac{d {\bf S}_{2}}{dt} & = & \frac{\delta_{2} }{r^3} {\bf L}_N \times {\bf S}_{2} \,.
\end{eqnarray}
Here $M = m_1 + m_2$ is the total mass, $\mu = m_1 m_2/M$ is the reduced mass, ${\bf L}_N = \mu {\bf r} \times
{\bf v}$ is the Newtonian contribution to the orbital angular momentum and 
\begin{equation}\label{Seff}
{\bf S_{\rm eff}} = \delta_1 {\bf S}_1 +\delta_2 {\bf S}_2
\end{equation}
with
\begin{eqnarray}
\delta_{1} &
= & 2 \left( 1 + \frac{3 m_2}{4 m_1} \right) \,, \\
\delta_{2} &
= & 2 \left( 1 + \frac{3 m_1}{4 m_2} \right) \, .
\end{eqnarray}
There are five constants of the motion: the magnitude of the angular
momentum $L$, the individual spin magnitudes $S_1,S_2$, the quantity
${\bf L}\cdot {\bf S}_{\rm eff}$, and the energy
\begin{equation}\label{energy}
E = \frac{{\bf p^2}}{2 \mu}-\frac{\mu M}{r}+\frac{ \bf L \cdot \bf S_{\rm eff}}{r^3}.
\end{equation}
The angular momentum has post-Newtonian and spin corrections: ${\bf L} = {\bf L}_N + {\bf L}_{PN}+
{\bf L}_{SO}+\dots$, which to 1.5 PN order are given by
\begin{eqnarray}
{\bf L}_{PN} &=& {\bf L}_N\left(\frac{1}{2} v^2(1-3\eta)+\frac{M}{r}(1+3\eta)\right) \, , \\
{\bf L}_{SO} &=& \frac{\mu}{r^3} {\bf r} \times ({\bf r} \times {\bf S}_{\rm eff}) \, ,
\end{eqnarray}
with $\eta = \mu/M$.
The momentum and velocity can be decomposed into radial and angular contributions:
\begin{eqnarray}
{\bf p} &=& p_r {\bf n} + p_\perp {\bf m} \nonumber \\
{\bf v} &=& v_r {\bf n} + v_\perp {\bf m} \, ,
\end{eqnarray}
where ${\bf n}= {\bf r}/r$, ${\bf m} = \hat{\bf L}_N \times {\bf n}$, $v_r = \dot{r}$ and
$p_\perp=L/r$. Combining the radial component of (\ref{reqn}), $p_r = \mu {\dot r}$,
with (\ref{energy}) yields
\begin{equation}
{\dot r}^2 =\frac{2E}{\mu}+\frac{2M}{r} - \frac{L^2}{\mu^2 r^2}
 -\frac{2 {\bf L}\cdot {\bf S_{eff}}}{\mu r^3} \, .
\end{equation}
Squaring (\ref{reqn}) and dropping higher order terms yields
\begin{equation}
v^2={\bf \dot r}\cdot {\bf \dot r} = \frac{p^2}{\mu^2} + \frac{2  \bf L \cdot \bf S_{\rm eff}}{\mu r^3}\, ,
\end{equation}
from which it follows that
\begin{equation}
v_\perp^2 = \frac{L^2}{\mu^2 r^2} +\frac{2 {\bf L}\cdot {\bf S_{eff}}}{\mu r^3} \, .
\end{equation}

Solving the above set of equations is complicated by the precession of the orbital plane, which causes
$v_\perp$ to appear as a mix of azimuthal and longitudinal motion. The equations are more
readily solved by transforming to a non-inertial frame that follows the precession of the
orbital plane. The precessing frame is defined by the condition
\begin{equation}
\frac{d \hat{\bf L}_N}{dt}\Big|_{\rm pre} = {\bf 0} = \frac{d \hat{\bf L}_N}{dt}
-{\boldsymbol \omega} \times \hat{\bf L}_N \, ,
\end{equation}
which implies that
\begin{equation}\label{omega}
{\boldsymbol \omega} = \frac{ {\bf S}_{\rm eff}}{r^3} \, .
\end{equation}
The velocity in the precessing frame is given by
\begin{eqnarray}\label{prec}
\frac{d{\bf r}}{dt}\Big|_{\rm pre} &=& \frac{d{\bf r}}{dt}-{\boldsymbol \omega} \times {\bf r } \, , \\
& = & \frac{{\bf p}}{\mu} \, ,
\end{eqnarray}
from which it follows that the orbital plane remains fixed in the precessing frame, ${\bf L}_N\cdot
\dot {\bf r}_{\rm pre} = 0$, and the radial motion is unchanged, ${\dot r}_{\rm pre}  = {\dot r}$.
Introducing the azimuthal coordinate $\phi$
in the orbital plane we have $v^2_{\rm pre} = {\dot r}^2 + r^2 {\dot \phi}^2$ and
\begin{equation}
{\dot \phi}^2 = \frac{L^2}{\mu^2 r^4}\, .
\end{equation}
The equations for ${\dot r}$ and ${\dot \phi}$ can be solved by introducing the generalized
Keplerian parameterization of the orbits~\cite{Damour:1985to}:
\begin{eqnarray}\label{eom}
{\bf r} &=& r\cos \phi \, {\bf p} + r\sin \phi \, {\bf l} \, \\
n t &=& u - e_t \sin u \, \\
r &=& a(1-e_r \cos u) \, \\
\phi &=& 2(k+1) \tan^{-1}\left[\left(\frac{1+e_{\phi}}{1-e_{\phi}}\right)^{1/2} \tan \frac{u}{2} \right] \,
\end{eqnarray}
where $u$ is the eccentric anomaly, $n=2\pi f$ is the mean motion with orbital frequency $f$, $k$ is the
fractional periastron advance per orbit, $a$ is the semi-major axis, and the regular Keplerian
eccentricity has split into the triad of time $e_t$, radial $e_r$, and angular $e_{\phi}$, eccentricities .
Here we have used a coordinate system defined by the Newtonian angular momentum $\hat{\bf L}_N$ and
the line of sight vector to the source $\hat{\bf N}$:
\begin{eqnarray}\label{basis}
{\bf p} &=& \hat{{\bf L}}_N \times \hat{{\bf N}}/\vert \hat{{\bf L}}_N \times \hat{{\bf N}} \vert \, , \nonumber \\
{\bf l} &=& \hat{{\bf L}_N} \times {\bf p} \, .
\end{eqnarray}
Following Ref.~\cite{Damour:1985to} and restoring the 1 PN contributions, we find
\begin{eqnarray}\label{kepler}
a &=& -\frac{\mu M}{2E}\left(1+\frac{1}{2}\left(7-\eta \right)\frac{E}{\mu}  -\frac{2\eta{\bf L \cdot S_{eff}} }{L^2}\frac{E}{\mu} \right) \, \\
e_r^2 &=& 1+2\frac{EL^2}{\mu^3 M^2}+\frac{E}{\mu}\left[2(\eta -6)+5(\eta -3)\frac{EL^2}{\mu^3 M^2}\right]\nonumber \\
&& +8\left(1+\frac{EL^2}{\mu^3 M^2}\right)\frac{\eta{\bf L \cdot S_{eff}} }{L^2}\frac{E}{\mu} \, \\
n & = & \frac{1}{M}\left(-\frac{2E}{\mu}\right)^{3/2}\left(1+\frac{1}{4}\left(15-\eta\right)\frac{E}{\mu}\right) \, \\
e_t &= & e_r\left(1+\left(8-3\eta\right)\frac{E}{\mu}- \frac{2\eta{\bf L \cdot S_{eff}} }{L^2}\frac{E}{\mu}\right) \, \\
k & = & \frac{3\mu^2 M^2}{L^2}\left(1- \frac{ \eta {\bf L \cdot S_{eff}} }{L^2} \right) \, \\
e_{\phi}&= & e_r \left(1-\frac{E}{\mu}\left(\eta- \frac{ 2\eta {\bf L \cdot S_{eff}} }{L^2} \right)\right) \, .
\end{eqnarray}
For completeness we have included the 1.5 PN correction to the perihelion precession, $k$, even though
it is formally a 2.5 PN order term.

The next step is to solve the spin-orbit precession equations (\ref{so}) to establish the time-dependent
transformation between the inertial and precessing frames of reference. We begin by writing ${\bf L}_N =
{\bar {\bf L}} + \delta {\bf L}$ where ${\bar {\bf L}}$ denotes the slowly changing, orbit averaged
angular momentum, and $\delta {\bf L}$ is a small periodic correction that varies on the orbital timescale.
Adopting a similar decomposition for the two spins we find
\begin{equation}
\delta {\bf L} = g(t) \left(\frac{M}{a}\right)^{3/2} 
                 \frac{\left({\bar {\bf S}}_{\rm eff} \times {\bar {\bf L}}\right)}{M^2} \, ,
\end{equation}
where
\begin{equation}
g(t) = \frac{\phi-u}{(1-e^2)^{3/2}}
-\frac{e\sin u}{(1-e^2)}\left(\frac{1}{\sqrt{1-e^2}}-\frac{1}{1-e\cos u}\right) \, .
\end{equation}
In the above expression it is understood that we are using the Newtonian limit for $u$, $\phi$, $e$ and $a$.
Note that the function $g(t)$ is periodic with period $T=1/f$. The $\delta {\bf L}$ term causes a periodic
variation in the observed waveforms of the same order as the 1.5 PN amplitude corrections discussed below.
The slowly varying orbit averaged expressions for the spins and angular momentum are found by numerically
integrating the coupled set of differential equations
\begin{eqnarray}\label{so_slow}
\frac{d {\bar {\bf L}}}{dt} & =& \frac{{\bf {\bar S}_{\rm eff}} \times {\bar {\bf L}}}{ a^3(1-e^2)^{3/2}}  \, ,\\
\frac{d {\bar {\bf S}}_{1}}{dt} & = & \frac{\delta_{1} }{a^3(1-e^2)^{3/2}} 
{\bar {\bf L}} \times {\bar {\bf S}}_{1} \,, \\
\frac{d {\bar {\bf S}}_{2}}{dt} & = & \frac{\delta_{2} }{a^3(1-e^2)^{3/2}} {\bar {\bf L}} 
\times {\bar {\bf S}}_{2} \,.
\end{eqnarray}
By solving the fast varying contribution to the precession equations analytically we have reduced the
computational cost by a factor of $(M/a)^{3/2}$ relative to solving the full equations. This completes
our solution of the dissipationless motion.

It is interesting to compare our solution to other expressions in the literature. Our expressions for
the quantities that enter the radial motion, $a, e_r, e_t$ and $n$, agree with those found by
Wex~\cite{wex}, Konigsdorffer and Gopakumar~\cite{Konigsdorffer:2005sc} and  Keresztes, Mik\'oczi
and Gergely~\cite{Keresztes:2005tp}. Konigsdorffer and Gopakumar~\cite{Konigsdorffer:2005sc}
provide expressions for higher PN order corrections that do not involve spin, while
Keresztes, Mik\'oczi and Gergely~\cite{Keresztes:2005tp} include the leading order spin-spin corrections.
For the angular motion Wex assumed that the spin precession could be neglected, leading to incorrect
expressions for $k$ and $e_\phi$. Konigsdorffer and Gopakumar included the effects of spin precession, but their
analysis was limited to the special case of simple precession, where either one spin vanishes or
the bodies have equal mass. Our expressions for $k$ and $e_\phi$ agree with theirs in the simple precession limit,
save for some spurious terms in their expressions that fail to vanish when the spins are set equal to zero.
Keresztes, Mik\'oczi and Gergely did not provide a solution for the angular motion.
More recently, Tessmer~\cite{Tessmer:2009yx} has provided a solution for general spin orientations
in the circular limit, but in a form that makes it difficult to compare to our solutions.

\section{Dissipation}

Dissipational effects first enter the equations of motion at 2.5 PN order. Rather than directly integrating
these equations, we adiabatically evolve the system by incrementing the energy and angular momentum according
to the flux equations. Because the dissipation occurs on a much longer timescale than the orbital or precession
motion, we begin by orbit averaging the instantaneous flux equations. Junker and Sch\"afer~\cite{Junker} carried out
the calculation to 1 PN order, and we now extend their calculation to include the spin-orbit effects at
1.5 PN order. The instantaneous expressions for the 1.5 PN fluxes were computed by
Kidder~\cite{Kidder:1995zr} using the covariant
spin supplementary condition, and by Zeng and Will~\cite{Zeng:2007bq} using the PNW spin supplementary
condition. We use the
latter expressions to be consistent with the choice we made for the dissipationless equations of motion.

The orbit-averaged flux at 1.5 PN order has two contributions: one from averaging the
0 PN flux over a 1.5 PN order orbit, $\langle F_0 \rangle_{1.5}$; the another from averaging
the 1.5 PN flux over a 0 PN order orbit, $\langle F_{1.5} \rangle_{0}$. For the energy these
are:
\begin{eqnarray}\label{E1}
\langle {\dot E}_0 \rangle_{1.5} &=& \frac{M^2 \mu}{15 a^7 (1-e^2)^{11/2}} 
\left[{\bf L} \cdot {\bf S}_{\rm eff} (96+276e^2\right. \nonumber \\
&& \left. +471e^4+74e^6) \right]
\end{eqnarray}
and
\begin{eqnarray}\label{E2}
\langle\dot{E}_{1.5}\rangle_0&=&\frac{M^2 \mu}{30 a^7 (1-e^2)^{11/2}} 
 \left[{\bf L \cdot S}  (784+5480e^2  \right. \nonumber\\
 && +3810e^4+195e^6)
 + {\bf L }\cdot{\bf Z} (432+2928e^2 \nonumber\\
 && \left. +1962e^4+96e^6)\right] \, .
\end{eqnarray}	
Here we have introduced the spin combinations
\begin{eqnarray}
{\bf S} &=& {\bf S}_1 + {\bf S}_2 \, \\
{\bf Z} &=& \frac{m_2}{m_1}{\bf S}_1 + \frac{m_1}{m_2}{\bf S}_2 \, ,
\end{eqnarray}
which are related to ${\bf S}_{\rm eff}$ by
\begin{equation}
{\bf S}_{\rm eff} = 2{\bf S} + \frac{3}{2} {\bf Z} \, .
\end{equation}
The contributions to the decay of the angular momentum are:
\begin{equation}\label{L1}
\langle {\dot{L}}_{0}\rangle_{1.5} = \frac{2 M^2 \mu^2 {\bf \hat L}\cdot {\bf S}_{\rm eff}
(16+33e^2+26e^4)}{5 a^5 (1-e^2)^{7/2}} 
\end{equation}
and
\begin{eqnarray}\label{L2}
 \langle{\bf \dot{L}}_{1.5}\rangle_0 &=&\frac{M^2 \mu^2}{15 a^5 (1-e^2)^{7/2}} \biggl \{ 
{\bf Z_\parallel} (144+592e^2+144e^4) \nonumber \\
&& + {\bf S_\parallel}(296+1032e^2+237e^4) \nonumber \\
&& - \frac{{\bf Z_\perp}}{4}(480+2496e^2+671e^4) \nonumber\\
&& - \frac{{\bf S_\perp}}{2}(332+1572e^2+435e^4)  \biggr \}
\end{eqnarray}	
where ${\bf S_\parallel} = {\bf \hat L}( {\bf \hat L} \cdot \bf S)$, ${\bf S_\perp} = \bf S - {\bf S_\parallel}$,
and similarly for ${\bf Z_\parallel}$ and  ${\bf Z_\perp}$. The terms parallel to the orbital angular
momentum contribute to the decay of the orbit (since ${\dot L} = {\bf \hat L} \cdot {\bf \dot{L}}$). The
orthogonal terms introduce additional, higher order precessional effects. Rieth and
Sch\"{a}fer~\cite{Rieth:1997mk} found expressions for the spin-orbit corrections to the energy and
angular momentum loss using a clever trick that allowed them to solve for the orbit averaged fluxes without
having to solve the equations of motion. They quote their expressions in terms of $E$ and $L$, but we
can invert (\ref{kepler}) to find expressions for $E$ and $L$ in terms of $a$ and $e_r$:
\begin{eqnarray}\label{EL}
E &=& -\frac{\mu}{2}\left(\frac{M}{a}\right)\left[
1-\frac{7-\eta}{4} \left(\frac{M}{a}\right) \right. \nonumber \\
&& \left. +\frac{{\bf \hat L} \cdot {\bf S_{eff}}}{M^2 (1-e_r^2)^{1/2}}\left(\frac{M}{a}\right)^{3/2}\right]  \\
L &=& \mu M \sqrt{1-e_r^2}\sqrt{\frac{a}{M}}\left[1 + \frac{(4+2e_r^2-\eta e_r^2)}{2(1-e_r^2)}\left(\frac{M}{a}\right) \right. \nonumber \\
&& \left. -\frac{{\bf \hat L} \cdot {\bf S_{eff}}}{M^2}\frac{(3+e_r^2)}{2(1-e_r^2)^{3/2}} \left(\frac{M}{a}\right)^{3/2}\right],
\end{eqnarray}
and after substituting these expressions into their equation (61) and converting the result into an energy flux by
dividing through by $-\mu$, we get perfect agreement with the sum of the terms in our equations (\ref{E1})
and (\ref{E2}). Similarly, after correcting their equation (70) by restoring a missing factor of $1/L^4$ in the
second term, we get perfect agreement with our expression for the rate of change of the angular momentum
from equations (\ref{L1}) and (\ref{L2}). It is worth noting that the expressions for the orbit-averaged energy and
angular momentum loss quoted by Rieth and Sch\"{a}fer~\cite{Rieth:1997mk} are gauge invariant
(i.e. they do not depend on the choice of SSC), while
our expressions are not gauge invariant since they involve non-invariant quantities such as $a$ and $e_r$.

The adiabatic decay of the orbits is found by applying the chain rule to our expressions (\ref{kepler}) for the
semi-major axis and radial eccentricity:
\begin{equation}
\langle \dot e_r \rangle = \left[ \frac{\partial e_r^2}{\partial E} \langle \dot E \rangle +
\frac{\partial e_r^2}{\partial L} \langle \dot L \rangle \right]/(2 e_r) \, .
\end{equation}
and similarly for $a$. The eccentricity evolution provides a good cross check for our calculations
as the collection of terms
in square brackets have to cancel to order $e_r^2$ to avoid unphysical behavior in the circular limit.

Putting everything together and restoring the 0 PN and 1 PN terms we have
\begin{eqnarray}\label{edotav}
\langle\dot{e_r}\rangle &=&-\frac{1}{15}\frac{\mu}{M^2}\left(\frac{M}{a}\right)^4\frac{e_r}{(1-e_r^2)^{5/2}}
\biggl \{  (304+121e_r^2)  \nonumber \\
&& -\left(\frac{M}{a}\right)\frac{1}{56 (1-e_r^2)}\left[8(16705+4676\eta) \right. \nonumber\\
&& \left. 12(9082+2807\eta)e_r^2-(25211-3388\eta)e_r^4 \right]   \nonumber\\
&& -\left(\frac{M}{a}\right)^{3/2}\frac{1}{2M^2(1-e_r^2)^{3/2}}\left[ (7032\, 
{\bf \hat L}\cdot{\bf S} \right. \nonumber \\
&& + 4408\, {\bf \hat L}\cdot{\bf Z}) 
+ (5592\, {\bf \hat L}\cdot{\bf S} + 2886\, {\bf \hat L}\cdot{\bf Z})e_r^2 \nonumber \\
&& \left. + (1313\,{\bf \hat L}\cdot{\bf S}+875\,{\bf \hat L}\cdot{\bf Z})e_r^4 \right] \biggl \} \, ,
\end{eqnarray}
and
\begin{eqnarray}\label{adotav}
\langle\dot{a}\rangle&=&-\frac{1}{15}\left(\frac{M}{a}\right)^3\frac{\eta}{(1-e_r^2)^{7/2}}
\biggl \{ 2(96+292e_r^2+37e_r^4) \nonumber \\
&& +\left( \frac{M}{a}\right) \frac{1}{14(1-e_r^2)} [ (14008+4704\eta)  \nonumber \\
&& +(80124 +21560\eta)e_r^2  
+(17325+10458\eta)e_r^4 \nonumber \\
&&-\frac{1}{2}(5501  -1036\eta)e_r^6] ]  \nonumber \\
&& +\left(\frac{M}{a}\right)^{3/2}\frac{1}{M^2 (1-e_r^2)^{3/2}}[ (2128\, {\bf \hat L}\cdot{\bf S}
+1440\, {\bf \hat L}\cdot{\bf Z}) \nonumber \\
&& + (7936 {\bf \hat L}\cdot{\bf S}
+4770 {\bf \hat L}\cdot{\bf Z})e_r^2 \nonumber\\
&&+ (3510\, {\bf \hat L}\cdot{\bf S} + 1737\, {\bf \hat L}\cdot{\bf Z})e_r^4 \nonumber \\
&& + (363\, {\bf \hat L}\cdot{\bf S}+222\, {\bf \hat L}\cdot{\bf Z})e_r^6 ]  \biggl \} \, .
\end{eqnarray} 
Gergely, Perj\'es and Vas\'uth~\cite{Gergely:1998sr} have performed a similar calculation using the
covariant SSC, and while it is generally impossible to compare results that use different
choices for the SSC, we note that the terms involving ${\bf \hat L}\cdot{\bf S}$ are in agreement,
while those involving ${\bf \hat L}\cdot{\bf Z}$ are not. The explantion is that the ${\bf \hat L}\cdot{\bf S}$
terms in the equations of motion are the same in the two SSC frames, while the ${\bf \hat L}\cdot{\bf Z}$
terms differ.

Our solution to the equations of motion is completed by using (\ref{EL}) to recast the other
Keplerian parameters as functions of $a$ and $e_r$:
\begin{eqnarray}
n&=&\frac{1}{M}\left(\frac{M}{a}\right)^{3/2}\left(1-
\left(\frac{M}{a}\right)\frac{\left(9-\eta\right)}{2} \right. \nonumber \\
&& \left. +\left(\frac{M}{a}\right)^{3/2}\frac{3 {\bf \hat L} 
\cdot {\bf S_{eff}}}{2 M^2 \sqrt{1-e^2_r}}\right) \, \\
e_t&=&e_r\left(1+\left(\frac{M}{a}\right)\frac{\left(3\eta-8\right)}{2}  \right. \nonumber \\
&& \left. +\left(\frac{M}{a}\right)^{3/2}\frac{{\bf \hat L \cdot S_{eff}} }{M^2 \sqrt{1-e^2_r}}\right) \, \\
k&=&\frac{3}{(1-e^2_r)}\left(\frac{M}{a}\right) \, \\
e_{\phi}&=&e_r \left(1+\frac{\eta}{2}\left(\frac{M}{a}\right)
-\left(\frac{M}{a}\right)^{3/2}\frac{{\bf \hat L \cdot S_{eff}} }{M^2 \sqrt{1-e^2_r}}\right) \,
\end{eqnarray}

\section{Waveforms}

With the orbital motion in hand, the final step is to derive expressions for the polarization
states of the gravitational waves. These can be written as
\begin{eqnarray}
h_+ &=& \frac{1}{2}(p_i p_j - q_i q_j)h^{ij}_{\rm TT} \nonumber \\
h_\times &=& \frac{1}{2}(p_i q_j + q_i p_j)h^{ij}_{\rm TT}
\end{eqnarray}
where $h^{ij}_{\rm TT}$ are the transverse-traceless components of the metric perturbation
and ${\bf p}$ and ${\bf q}$ are unit vectors orthogonal to the line of sight vector
$\hat{ {\bf N}}$ defined by
\begin{eqnarray}\label{pol}
{\bf p} &=& \hat{{\bf L}}_N \times \hat{{\bf N}}/\vert \hat{{\bf L}}_N \times \hat{{\bf N}} \vert \, , \nonumber \\
{\bf q} &=& \hat{{\bf N}} \times {\bf p} \, .
\end{eqnarray}
Expressions for $h^{ij}_{\rm TT}$ out to the
requisite order have been computed by
Kidder~\cite{Kidder:1995zr} using the covariant spin-supplementary condition. To our knowledge,
a similar calculation has not been done using the PNW spin supplementary condition, but
transforming Kidder's expressions to the PNW frame reveals no change to 1.5 PN order,
which is all we need for our current purposes (the first changes appear at 2.5 PN order).

Maj\'{a}r and Vas\'{u}th~\cite{Majar:2008zz} have provided formal expressions for the instantaneous
polarization states
using a corotating coordinate system. They first define an ``invariant'' source frame using the line of
sight vector $\hat{ {\bf N}}$ and the total angular momentum vector ${\bf J} = {\bf L} + {\bf S}$, which
they then relate to a corotating system attached to the orbital angular momentum ${\bf L}_N$ and
orbital separation vector ${\bf r}$. When dissipation is included, ${\bf J}$ slowly evolves and it
is necessary to introduce a new invariant coordinate system. A convenient choice might be a frame
related to the detectors, such as the Barycenter or Geocenter frames, which are
approximately invariant on the observational timescale of a space based or ground based detector.
Thus our description of the waveforms involves three reference frames: the detection frame; the
source frame; and the orbital frame. The rotation between the source frame and orbital frame occurs on
the orbital timescale, while the rotation between the source frame and the detection frame occurs on
the more sedate dissipation timescale.

The source frame is described by the triad $\{ \hat{{\bf i}}, \hat{{\bf j}}, \hat{{\bf k}} \}$, where 
$\hat{{\bf k}}=\hat{{\bf J}}$, $\hat{{\bf i}}=\hat{{\bf j}}\times \hat{{\bf J}}$,
$\hat{{\bf j}}=\hat{{\bf J}}\times \hat{{\bf N}}/\sin\gamma$, and $\cos\gamma = \hat{{\bf J}}\cdot\hat{{\bf N}}$. 
The comoving orbital frame is described by the triad $\{ \hat{{\bf x}}, \hat{{\bf y}}, \hat{{\bf z}} \}$, where
$\hat{{\bf x}} = \hat{{\bf r}}$, $\hat{{\bf z}} = \hat{{\bf L}}_N$ and
$\hat{{\bf y}} = \hat{{\bf L}}_N \times \hat{{\bf r}}$. 
The corotating system is related to the source frame by three time dependent Euler angles $\{\Phi, i_S, \Psi\}$,
where $\Psi$ is the orbital phase in the source frame, $\Phi$ is the precession angle of the orbital plane,
and $i_S$ is the precession cone angle. Applying these rotations to $\hat{ {\bf N}}$, we find it has the
following components in the corotating frame:
\begin{eqnarray}
N_x &=&  \cos{\Psi} \cos{\Phi} \sin{\gamma} - \sin{\Psi}(\sin{\Phi} \sin{\gamma} \cos{i_S} \nonumber \\
&& - \cos{\gamma} \sin{i_S}) \nonumber \\
N_y &=& -\sin{\Psi} \cos{\Phi} \sin{\gamma} - \cos{\Psi}(\sin{\Phi} \sin{\gamma} \cos{i_S} \nonumber \\
&& - \cos{\gamma} \sin{i_S})  \nonumber \\
N_z &=& \sin{i_S} \sin{\Phi} \sin{\gamma}  + \cos{\gamma} \cos{i_S} \, .
\end{eqnarray}
The components of ${\bf p}$ and ${\bf q}$ in the corotating frame are then
\begin{eqnarray}
p_i &=& -N_y/\sqrt{N_x^2+N_y^2}  \nonumber \\
p_j &=& N_x/\sqrt{N_x^2+N_y^2} \nonumber \\
p_k &=&  0 \, ,
\end{eqnarray}
and
\begin{eqnarray}
q_i &=& -N_x N_z/\sqrt{N_x^2+N_y^2}  \nonumber \\
q_j &=& N_y N_z /\sqrt{N_x^2+N_y^2} \nonumber \\
q_k &=& \sqrt{N_x^2+N_y^2} \, .
\end{eqnarray}
The mapping between the detection frame and the source frame can be accomplished by evolving
the components of the spins and angular momentum in the detection frame, then solving for
the angles that appear in the corotating frame using the relations
\begin{eqnarray}
\cos{i_{S}} &=&  {\bf \hat J} \cdot {\bf \hat L}_N  \\
\cos{\gamma} &=& {\bf \hat J} \cdot {\bf \hat N}   \\
\cos{\Phi} & =& \frac{\left({\bf \hat N} \times {\bf \hat J}\right) \cdot \left( {\bf \hat J} \times {\bf \hat L}_N \right)}{\vert{\bf \hat N} \times {\bf \hat J}\vert \vert {\bf \hat J} \times {\bf \hat L}_N\vert} \, ,
\end{eqnarray}
and~\cite{Arun:2008kb}
\begin{eqnarray}
\dot{\Psi} &=& \dot{\phi} - \cos{i_S}\,\dot{\Phi} \, .
\end{eqnarray}

For completeness, and also to correct several errors in the original paper, we re-compute and quote the
explicit expression for the polarization states out to 1.5 PN order. The 1.5 PN terms are separated
into spin-orbit (SO) and other contributions.
\begin{widetext}
\begin{eqnarray}\label{hplus}
h_+^0&=&\left(\dot{r}^2-\frac{M}{r}\right)(p_x^2-q_x^2)+2v_{\perp}\dot{r}(p_xp_y-q_xq_y)
+ v_{\perp}^2(p_y^2-q_y^2)\ ,\\
h_+^{0.5}&=&\frac{\delta m}{M}\left(\left(\dot{r}\left[\frac{2M}{r}-\dot{r}^2\right]N_x+ v_{\perp}\left[\frac{M}{2r}-\dot{r}^2\right]N_y\right)(p_x^2-q_x^2)\right.\nonumber\\ 
&&+v_{\perp}\left(\left[\frac{3M}{r}-2\dot{r}^2\right]N_x-2v_{\perp}\dot{r}N_y\right)(p_xp_y-q_xq_y)\nonumber\\
&&-\left.v_{\perp}^2\left(\dot{r}N_x+v_{\perp}N_y\right)(p_y^2-q_y^2)\right)\ ,\\
h_+^1&=&\frac{1}{6}\biggl((1-3\eta)\biggl[\left(-\frac{21\dot{r}^2M}{r}+\frac{3Mv^2}{r}+6\dot{r}^4 +\frac{7M^2}{r^2}\right)N_x^2 \nonumber\\
&&+ 4v_{\perp}\dot{r}\left(-\frac{6M}{r}+3\dot{r}^2\right)N_xN_y +2v_{\perp}^2\left(3\dot{r}^2-\frac{M}{r}\right)N_y^2\biggr] \nonumber\\
&&+ \left[\frac{(19+9\eta)\dot{r}^2M}{r}+(3-9\eta)v^2\dot{r}^2-\frac{(10+3\eta)v^2M}{r}+\frac{29M^2}{r^2}\right]\biggr)(p_x^2-q_x^2)\nonumber\\
&&+\frac{v_{\perp}}{6}\left((1-3\eta)\left[6\dot{r}\left(-\frac{5M}{r}+2\dot{r}^2\right)N_x^2 +8v_{\perp}\left(-4\frac{M}{r}+3\dot{r}^2\right)N_xN_y+ 12v_{\perp}^2\dot{r}N_y^2\right]\right.\nonumber\\
&&+\left.6\dot{r}\left[\frac{(2+4\eta)M}{r}+(1-3\eta)v^2\right]\right)(p_xp_y-q_xq_y)\nonumber\\ 
&&+\frac{v_{\perp}^2}{6}\left((1-3\eta)\left[2\left(-\frac{7M}{r}+3\dot{r}^2\right)N_x^2+12v_{\perp}\dot{r}N_xN_y+6v_{\perp}^2N_y^2\right]\right.\nonumber\\
&&+\left.\left[-\frac{(4-6\eta)M}{r}+ (3-9\eta)v^2\right]\right)(p_y^2-q_y^2)\ ,\\
h_+^{SO}&=&-\frac{1}{r^2}\left[({\bf \Delta{\mbox{\boldmath$\cdot$}}q})p_x
+({\bf \Delta{\mbox{\boldmath$\cdot$}}p})q_x\right]\ ,\nonumber\\
\end{eqnarray}
\begin{eqnarray}
h_+^{1.5}&=&\frac{\delta m}{M}\left\{(1-2\eta)\left(\dot{r}\left[\frac{3\dot{r}^2M}{4r}
-\frac{v^2M}{r}- \frac{41M^2}{12r^2}-\dot{r}^4\right]N_x^3\right.\right.\nonumber\\
&&+\left.v_{\perp}\left[\frac{85\dot{r}^2M}{8r}-\frac{9v^2M}{8r}
- \frac{7M^2}{2r^2}-3\dot{r}^4\right]N_x^2N_y\right.\nonumber\\
&&+\left.3\dot{r}v_{\perp}^2\left[\frac{2M}{r}- \dot{r}^2\right]N_xN_y^2
+v_{\perp}^3\left[\frac{M}{4r}-\dot{r}^2\right]N_y^3\right)\nonumber\\
&&+\dot{r}\left[-\frac{(10+7\eta)\dot{r}^2M}{2r}+\frac{(2+\eta)v^2M}{2r}
-\frac{(59-30\eta)M^2}{12r^2}-\frac{(1-5\eta)v^2\dot{r}^2}{2}\right]N_x\nonumber\\
&&+\left.v_{\perp}\left[-\frac{(25+26\eta)\dot{r}^2M}{8r}+\frac{(7-2\eta)v^2M}{8r}- \frac{(26-3\eta)M^2}{6r^2}
-\frac{(1-5\eta)v^2\dot{r}^2}{2}\right]N_y\right\}(p_x^2-q_x^2)\nonumber\\
&&+v_{\perp}\frac{\delta m}{M}\left\{(1-2\eta)\left(\left[\frac{\dot{r}^2M}{4r}-\frac{7v^2M}{4r}
-\frac{11M^2}{r^2}-2\dot{r}^4\right]N_x^3+v_{\perp}\dot{r}\left[\frac{16M}{r}
-6\dot{r}^2\right]N_x^2N_y\right.\right.\nonumber\\
&&+\left.3v_{\perp}^2\left[\frac{5M}{2r}-2\dot{r}^2\right]N_xN_y^2-2v_{\perp}^3\dot{r}N_y^3\right) \nonumber\\
&&+\left[-\frac{(49+14\eta)\dot{r}^2M}{4r}+\frac{(11-6\eta)v^2M}{4r}-\frac{(32-9\eta)M^2}{3r^2}
-(1-5\eta)v^2\dot{r}^2\right]N_x\nonumber\\
&&-\left.v_{\perp}\dot{r}\left[\frac{(2+6\eta)M}{r}+ (1-5\eta)v^2\right]N_y\right\}(p_xp_y-q_xq_y)\nonumber\\
&&+v_{\perp}^2\frac{\delta m}{M}\left\{(1-2\eta)\left(-\dot{r}\left[\frac{5M}{4r}+\dot{r}^2\right]N_x^3
+v_{\perp}\left[\frac{29M}{4r}-3\dot{r}^2\right]N_x^2N_y-3v_{\perp}^2\dot{r}N_xN_y^2
-v_{\perp}^3N_y^3\right)\right.\nonumber\\
&&-\left.\dot{r}\left[\frac{(7+3\eta)M}{r}+\frac{(1-5\eta)v^2}{2}\right]N_x+ v_{\perp}\left[\frac{(3-8\eta)M}{4r}
-\frac{(1-5\eta)v^2}{2}\right]N_y\right\}(p_y^2-q_y^2)\ ,\nonumber\\
\end{eqnarray}
\begin{eqnarray}\label{hcross}
h_{\times}^0&=&2\left(\left(\dot{r}^2-\frac{M}{r}\right)p_xq_x+v_{\perp}\dot{r}(p_xq_y+q_xp_y) +v_{\perp}^2p_yq_y\right)\ ,\\
h_{\times}^{0.5}&=&\frac{\delta m}{M}\left[\left(\dot{r}\left[\frac{4M}{r}-2\dot{r}^2\right]N_x+v_{\perp}\left[\frac{M}{r}
-2\dot{r}^2\right]N_y\right)p_xq_x\right.\nonumber\\ 
&&\left.+v_{\perp}\left(\left[\frac{3M}{r}-2\dot{r}^2\right]N_x-2v_{\perp}\dot{r}N_y\right)(p_xq_y+q_xp_y)\right.\nonumber\\
&&\left.-2v_{\perp}^2\left(\dot{r}N_x+v_{\perp}N_y\right)p_yq_y\biggr]\right.\ ,\\
h_{\times}^{1}&=&\frac{1}{3}\biggl((1-3\eta)\biggl(\left[-\frac{21\dot{r}^2M}{r}+\frac{3Mv^2}{r}+ 6\dot{r}^4 +\frac{7M^2}{r^2}\right]N_x^2\nonumber\\
&&+ 4v_{\perp}\dot{r}\left[-\frac{6M}{r}+3\dot{r}^2\right]N_xN_y+2v_{\perp}^2\left[3\dot{r}^2-\frac{M}{r}\right]N_y^2\biggr)\nonumber\\
&&+\left[\frac{(19-9\eta)\dot{r}^2M}{r}+(3-9\eta)v^2\dot{r}^2-\frac{(10+3\eta)v^2M}{r}+\frac{29M^2}{r^2}\right] \biggr) p_xq_x\nonumber\\
&&+\frac{v_{\perp}}{6}\left((1-3\eta)\left(6\dot{r}\left[-\frac{5M}{r}+2\dot{r}^2\right]N_x^2 
+8v_{\perp}\left[-4\frac{M}{r}+3\dot{r}^2\right]N_xN_y+ 12v_{\perp}^2\dot{r}N_y^2\right)\right.\nonumber\\
&&+\left.6\dot{r}\left[\frac{(2+4\eta)M}{r}+(1-3\eta)v^2\right]\right)(p_xq_y+q_xp_y)\nonumber\\ 
&&+\frac{v_{\perp}^2}{3}\biggl((1-3\eta)\biggl(2\left[-\frac{7M}{r}+3\dot{r}^2\right]N_x^2+12v_{\perp}\dot{r}N_xN_y+6v_{\perp}^2N_y^2\biggr)\nonumber\\
&&-\left[\frac{(4-6\eta)M}{r}-(3-9\eta)v^2\right]\biggr)p_yq_y\ ,\\
h_{\times}^{SO}&=&-\frac{1}{r^2}\left[({\bf\Delta{\mbox{\boldmath$\cdot$}}q})q_x
-({\bf\Delta{\mbox{\boldmath$\cdot$}}p})p_x\right]\ ,\nonumber\\
\end{eqnarray}
\begin{eqnarray}
h_{\times}^{1.5}&=&\frac{\delta m}{M}\left\{(1-2\eta)\left(\dot{r}\left[\frac{3\dot{r}^2M}{2r}-\frac{2v^2M}{r}
- \frac{41M^2}{6r^2}-2\dot{r}^4\right]N_x^3\right.\right.\nonumber\\
&&+\left.v_{\perp}\left[\frac{85\dot{r}^2M}{4r}-\frac{9v^2M}{4r}- \frac{7M^2}{r^2}-6\dot{r}^4\right]N_x^2N_y\right. \nonumber\\
&&+\left. 6\dot{r}v_{\perp}^2\left[\frac{2M}{r}-\dot{r}^2\right]N_xN_y^2
+v_{\perp}^3\left[\frac{M}{2r}-2\dot{r}^2\right]N_y^3\right)\nonumber\\
&&+\dot{r}\left[-\frac{(10+7\eta)\dot{r}^2M}{r}+\frac{(2+\eta)v^2M}{r}- \frac{(59-30\eta)M^2}{6r^2}
- (1-5\eta)v^2\dot{r}^2\right]N_x\nonumber\\
&&+\left.v_{\perp}\left[-\frac{(25+26\eta)\dot{r}^2M}{4r}+\frac{(7-2\eta)v^2M}{4r}
- \frac{(26-3\eta)M^2}{3r^2}- (1-5\eta)v^2\dot{r}^2\right]N_y\right\}p_xq_x\nonumber\\
&&+v_{\perp}\frac{\delta m}{M}\left\{(1-2\eta)\left(\left[\frac{\dot{r}^2M}{4r}-\frac{7v^2M}{4r}
- \frac{11M^2}{r^2}-2\dot{r}^4\right]N_x^3+v_{\perp}\dot{r}\left[\frac{16M}{r}
- 6\dot{r}^2\right]N_x^2N_y\right.\right.\nonumber\\
&&+\left.3v_{\perp}^2\left[\frac{5M}{2r}-2\dot{r}^2\right]N_xN_y^2-2v_{\perp}^3\dot{r}N_y^3\right)
+\left[-\frac{(49+14\eta)\dot{r}^2M}{4r}+\frac{(11-6\eta)v^2M}{4r}\right.\nonumber\\
&&-\left.\left.\frac{(32-9\eta)M^2}{3r^2}-(1-5\eta)v^2\dot{r}^2\right]N_x
- v_{\perp}\dot{r}\left[\frac{(2+6\eta)M}{r}+ (1-5\eta)v^2\right]N_y\right\}(p_xq_y+q_xp_y)\nonumber\\
&&+v_{\perp}^2\frac{\delta m}{M}\left\{(1-2\eta)\left(-\dot{r}\left[\frac{5M}{2r}+\dot{r}^2\right]N_x^3
+v_{\perp}\left[\frac{29M}{2r}-6\dot{r}^2\right]N_x^2N_y-6v_{\perp}^2\dot{r}N_xN_y^2
-2v_{\perp}^3N_y^3\right)\right.\nonumber\\
&&-\left.\dot{r}\left[\frac{(14+6\eta)M}{r}+(1-5\eta)v^2\right]N_x
+ v_{\perp}\left[\frac{(3-8\eta)M}{2r}-(1-5\eta)v^2\right]N_y\right\}p_yq_y\ ,\nonumber\\
\end{eqnarray}
\end{widetext}
where $\delta m=m_2-m_1$, and ${\bf\Delta}=M({{\bf S}_2}/{m_2}-{{\bf S}_1}/{m_1})$.

We have checked that the waveforms we get by substituting in our solution for the orbital motion and
spin-orbit precession are in {\em perfect} agreement with the Lang and Hughes waveforms~\cite{Lang:1900bz} in
the circular limit.

\section{Future Work}

It would be desirable to extend our treatment to higher post-Newtonian order.
The necessary building blocks are known to 2.5 PN order, and while there is no fundamental barrier
to going to higher order, there are some new effects to contend with. We will briefly
describe some of the issues that crop up at 2 PN order. At this order the precession equations read
\begin{eqnarray}
\frac{d {\bf S}_{1}}{dt} & = & \frac{\delta_{1} }{r^3} {\bf L}_N \times {\bf S}_{1}
+\frac{3}{r^3}\left( {\bf S}_1 \times {\bf S}_2 + ({\bf n}\cdot{\bf S}_2)
({\bf n}\times {\bf S}_1)\right),\nonumber \\
\frac{d {\bf S}_{2}}{dt} & = & \frac{\delta_{2} }{r^3} {\bf L}_N \times {\bf S}_{2} 
+\frac{3}{r^3}\left( {\bf S}_2 \times {\bf S}_1 + ({\bf n}\cdot{\bf S}_1)({\bf n}\times {\bf S}_2)\right),
\nonumber \\
\frac{d {\bf L}_N}{dt} & =& \frac{1}{ r^3} {\bf S_{\rm eff}} \times {\bf L}_N 
 -\frac{3}{r^3}\left( ({\bf n}\cdot{\bf S}_2)({\bf n}\times {\bf S}_1) \right. \nonumber \\
&& \hspace*{1.1in} \left. +({\bf n}\cdot{\bf S}_2)({\bf n}\times {\bf S}_1)\right),
\end{eqnarray}
from which it follows that ${\bf L}\cdot {\bf S}_{\rm eff}$ and the $L_N$ are no longer constant.
The condition $d \hat{\bf L}_N/dt\vert_{\rm pre}={\bf 0}$ demands that the precessing frame rotates
with angular velocity
\begin{equation}
{\boldsymbol \omega} = \frac{ {\bf S}_{\rm eff}}{r^3} -\frac{3{\bf n}}{r^3 L_N}\left(
({\bf n}\cdot {\bf S}_2)(\hat{\bf L}_N\cdot {\bf S}_1)+({\bf n}\cdot {\bf S}_1)(\hat{\bf L}_N\cdot {\bf S}_2)
\right) \, .
\end{equation}
Note that the new terms in ${\boldsymbol \omega}$ have no affect on the mapping between the
velocity in the inertial and precessing frames. As before, the equations for $\dot{r}$ and ${\dot \phi}$ 
can be written as polynomials in $1/r$, with new ${\rm spin}^2$ terms from the 2PN order Hamiltonian.
In contrast to what we found at 1.5 PN order, the spin-dependent coefficients in the polynomial
expansion are no longer constant. The time dependence of the coefficients prevents us from finding an
exact generalized Keplerian solution to the equations of motion in the precessing frame beyond 1.5 PN
order. On the other hand, we know that the spin dependent coefficients are approximately constant on
the orbital time scale since the time dependence enters at 2 PN order. Thus we can find a
solution for the Keplerian parameters by treating the spin dependent terms as constants, which
are then updated adiabatically via the precession equations, just as the total energy and angular momentum
are updated adiabatically via the dissipation equations.

As first noted by Damour and Sch\"{a}fer\cite{Damour1988}, and later solved by
Sch\"{a}fer and Wex~\cite{Schafer1993}, the generalized Keplerian
parameterization has to be further generalized to handle 2 PN terms in the equations of motion
for non-spinning bodies. In particular, they found it necessary to introduce terms involving the quantity
\begin{equation}
v = 2  \tan^{-1}\left[\left(\frac{1+e_{\phi}}{1-e_{\phi}}\right)^{1/2} \tan \frac{u}{2} \right] \, ,
\end{equation}
which is closely related to the orbital angle $\phi$. Some of the spin-spin terms that appear at 2 PN
order will require addition terms that depend on $v$. In particular, there are terms in the equation of
motion of the form $({\bf S}_1 \cdot {\bf n})({\bf S}_2 \cdot {\bf n})$, which produce terms like
$({\bf S}_1 \cdot {\bf p})({\bf S}_2 \cdot {\bf p})\cos^2 v$ in the generalized Keplerian parameterization.
In the adiabatic approach, spin dependent coefficients such as ${\bf S}_1 \cdot {\bf q}$,
${\bf S}_2 \cdot {\bf p}$ and ${\bf S}_1 \cdot {\bf S}_2$ are treated as constants when solving the
equations of motion. Expressions for the 2PN spin-spin corrections to the radial motion have already
been derived using the PNW spin supplementary condition~\cite{Keresztes:2005tp}. However,
given that the terms involving ${\bf L}\cdot {\bf S}_{\rm eff}$ are no longer constant
when 2 PN effects are taken into account, there is little advantage to using the PNW spin supplementary condition
beyond 1.5 PN order. Indeed, it may be wise to adopt the covariant spin supplementary condition since
this is the gauge in which the higher order corrections to the instantaneous waveforms have already been
derived. 

\section{Acknowlegements}
This work was supported by NASA grant NNX07AJ61G. We thank L. Gergely for bringing
Ref.~\cite{Rieth:1997mk} and Ref.~\cite{Gergely:1998sr} to our attention, and for helpful discussions
regarding dissipational effects under different choices for the spin supplementary condition.

\end{document}